\documentclass[useAMS,usenatbib]{mn2e}

\usepackage{graphicx, url}

\newcommand{\hi}{{\rm H{\textsc{i}}\,}}
\newcommand{\hii}{{\rm H{\textsc{ii}}\,}}

\newcommand{\hei}{{\rm He{\textsc{i}}\,}}
\newcommand{\heii}{{\rm He{\textsc{ii}}\,}}

\newcommand{\sulii}{{\rm S{\textsc{ii}}\,}}

\newcommand{\fevii}{{\rm Fe{\textsc{vii}}\,}}

\title[Electron Distribution in the Galactic Disk]{Electron Distribution in the Galactic Disk -
Results From a Non-Equilibrium Ionization Model of the ISM}
\author[M.A. de Avillez et al.]{M. A. de Avillez$^{1,3}$\thanks{E-mail: mavillez@galaxy.lca.uevora.pt} 
Ashish Asgekar$^{2}$, Dieter Breitschwerdt$^{3}$ and Emanuele Spitoni$^{1}$\\
$^{1}$Department of Mathematics, University of \'Evora, R. Romao Ramalho 59, 7000 \'Evora, Portugal\\
$^{2}$ASTRON, the Netherlands Institute for Radio Astronomy, P.O. Box 2, 7990 AA Dwingeloo, The Netherlands\\
$^{3}$Zentrum f\"ur Astronomie und Astrophysik, Technische Universit\"at Berlin, Hardenbergstrasse 36, D-10623
Berlin, Germany}
\begin{document}

\date{Accepted YYYY Month DD. Received YYYY Month DD; in original form 2011 December XX}

\pagerange{\pageref{firstpage}--\pageref{lastpage}} \pubyear{2011}

\maketitle

\label{firstpage}

\begin{abstract}
Using three-dimensional non-equilibrium ionization (NEI) hydrodynamical simulation of the interstellar
medium (ISM), we study the electron density, $n_{e}$, in the Galactic disk and compare it with
the values derived from dispersion measures towards pulsars with known distances located up to 200 pc on
either side of the Galactic midplane.

The simulation results, consistent with observations, can be summarized as follows: (i) the DMs in the
simulated disk lie between the maximum and minimum observed values, (ii) the $\log \langle n_e \rangle$
derived from lines of sight crossing the simulated disk follows a Gaussian distribution centered at $\mu=-1.4$
with a dispersion $\sigma=0.21$, thus, the Galactic midplane $\langle n_{e} \rangle=0.04\pm 0.01$ cm$^{-3}$,
(iii) the highest electron concentration by mass (up to 80\%) is in the thermally unstable regime
($200<\mbox{T}<10^{3.9}$ K), (iv) the volume occupation fraction of the warm ionized medium is 4.9-6\%, and
(v) the electrons have a clumpy distribution along the lines of sight.
\end{abstract}


\section{Introduction}

Dispersion measures (DMs) towards pulsars with known distance $d$ can be used to derive the mean
electron density in the Galaxy through the relation $\langle n_{e}\rangle=DM/d$. For pulsars
located at $\left|b\right|<5\degr$ the derived mean electron density in the Galactic plane is $\langle
n_{e}\rangle \sim 0.02-0.1$ cm$^{-3}$, and $0.01-0.017$ cm$^{-3}$ in the spiral arms and inter-arm regions,
respectively (Ferri\`ere 2001; Gaensler et al. 2008). Berkhuijsen \& Fletcher (2008), using the DMs
towards 34 pulsars, mostly outside the galactic plane, showed that the $\log \langle n_{e}\rangle$ follows a
Gaussian distribution, relating it to the nature of the interstellar turbulence (see e.g. reviews by Elmegreen
\& Scalo 2004). So far, numerical (magnetized and unmagnetized) 3D simulations have assumed collisional
ionization equilibrium (CIE) conditions for the ISM. In fact the thermal evolution of the ISM is determined by
heating and cooling processes, which in general are not synchronized with ionization and recombination
processes, respectively (e.g., Kafatos 1973; Shapiro \& Moore 1976). Hence, below $10^6$ K, deviations from
CIE conditions occur, thereby affecting the ionization structure of the interstellar gas, and thus the local
electron density. In particular, if delayed recombination plays a role, the number of free electrons may be
severely underestimated (Breitschwerdt \& Schmutzler 1994). 

Here we study the electron density, $n_{e}$, and it's mean, $\langle n_e \rangle$, in the Galactic disk (up to
200 pc on either side of the midplane) using the first to date global hydrodynamical simulation of the
interstellar gas, evolving under NEI conditions. Furthermore, we compare simulation results with estimates of
$\langle n_e \rangle$ obtained from available pulsar DMs. In forthcoming papers we explore the
$n_{e}$ distribution and it's topology in the thick disk and halo of the Milky Way and other
galaxies.

This paper is organized as follows: Section 2 deals with the model setup; Sections 3
and 4 present the observational data, and simulation results, respectively. A discussion and final
remarks in Section 5 close the paper.

\section{Model and Numerical Setup}

We simulate hydrodynamically the supernova-driven ISM in a patch of the Galaxy centered at the Solar circle
with an area of 1 kpc$^{2}$ and extending to 15 kpc on either side of the Galactic midplane following de
Avillez \& Breitschwerdt (2005, 2007; AB0507). The simulations are carried out with the EAF-parallel adaptive
mesh refinement code coupled to the newly developed E(A+M)PEC code\footnote{www.lca.uevora.pt/research.html
presents a description of the code, ionization fractions, cooling, and emission spectra.}; Avillez \&
Breitschwerdt 2012) featuring the time-dependent calculation on the spot (at each grid cell) of
the ionization structure of H, He, C, N, O, Ne, Mg, Si, S and Fe and emissivities. 

The physical model includes supernovae (SNe) types Ia, Ib+c, and II, a gravitational field provided by the
stars in the disk, local self-gravity (excluding the contribution from the newly formed stars), heat
conduction (Dalton \& Balbus 1993), uniform heating due to a UV radiation field normalized to the Galactic
value and varying with $z$, and photoelectric heating of grains and polycyclic aromatic hydrocarbons.

E(A+M)PEC uses the recommended abundances of Asplund et al. (2009), and calculates electron impact ionization,
inner-shell excitation auto-ionization, radiative and dielectronic recombination, charge-exchange reactions
(recombination with \hi~and \hei,~and ionization
with \hii and \heii), continuum (bremsstrahlung, free-bound, two-photon) and line (permitted, semi-forbidden
and forbidden) emission in the range 1\AA~-610$\mu$. The code also includes ionization of \hi by Lyman
continuum photons emitted during the recombination of helium. The internal energy of the plasma includes the
contributions due to the thermal translational energy plus the energy stored in (or delivered) from
high ionization stages. Electron impact ionization rates are taken from Dere (2007), while radiative
and dielectronic recombination rates are based on AUTOSTRUCTURE calculations (Badnell et al. 2003,
Badnell 2006a\footnote{amdpp.phys.strath.ac.uk/tamoc/DATA/}) including the latest corrections to Fe
ions by Badnell (2006b), Nikolic et al. (2010), and Schmidt et al. (2008). Radiative and
dielectronic recombination rates for \sulii and \fevii are from Mazzotta el al. (1998). For the
remaining ions we adopt the total recombination rates derived with the unified electron-ion
recombination method (Nahar \& Pradhan 1994) and available
at NORAD-Atomic-Data\footnote{www.astronomy.ohio-state.edu/$\sim$nahar and references therein.}.

A coarse grid resolution of 8 pc is used, while the finest AMR resolution is $0.5\,$pc (4 levels of
refinement) for $\left|z\right|\leq 2$ kpc, 4 pc for $\left|z\right|>4$ kpc, and 1 pc elsewhere.
Periodic and outflow boundary conditions are set along the vertical faces and top/bottom
($z=\pm 15$ kpc) of the grid, respectively.
\begin{figure}
\centering
\includegraphics[width=0.9\hsize,angle=0]{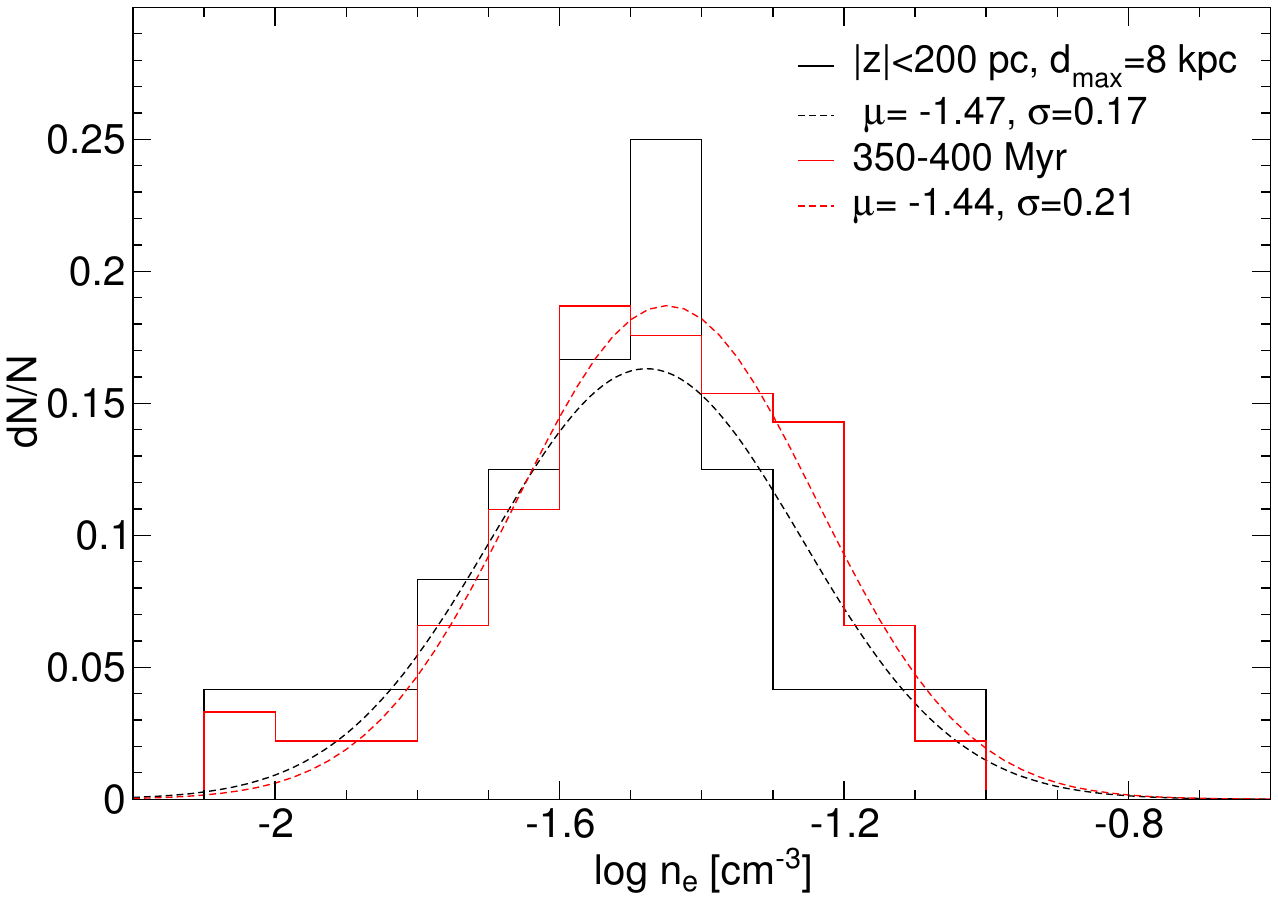}
\caption{Histogram (solid black line) and Gaussian fit (dashed black line) of $\log \langle
n_{e} \rangle$ distribution as obtained with a series of 24 pulsars located at $200 < \mbox{distance} <
8000$~pc from the Sun, with $\left|z\right|\leq 200$ pc. The mean value is $\log \langle n_{e}
\rangle = -1.47\pm0.02$ and $\sigma=0.17\pm0.02$. Red solid and dashed lines refer, respectively, to
the histogram and Gaussian fit (centered at $\mu=-1.44$ cm$^{-3}$ and having a dispersion
$\sigma=0.21$) of the $\langle n_{e}\rangle_{350-400\mbox{ Myr}}$ (\S4).}
\label{observ}
\end{figure}

\section{The Electron Density in the Disk}

To make a detailed comparison between simulation results (discussed below) and the electron density
distribution derived from DMs towards pulsars, we reassessed the existing data to select
pulsars with best possible distance estimates. Density PDFs were created using pulsars located up to 8 kpc
away from the Sun and with $|z|<200$ pc from the Galaxy's midplane.

\begin{figure*}
\centering
\includegraphics[width=0.45\hsize,angle=0]{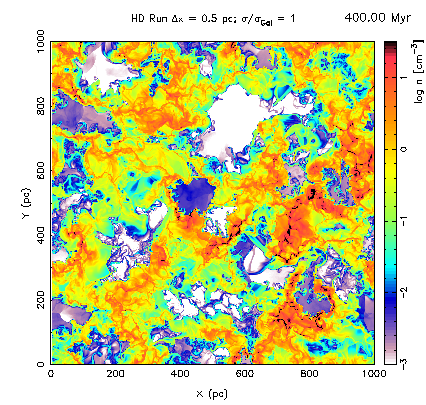}
\includegraphics[width=0.45\hsize,angle=0]{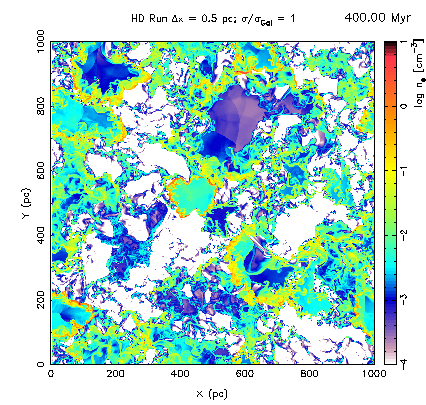}
\caption{Total (left) and electron (right) density distributions (in log scale) in the Galactic midplane at
400 Myr of evolution. Red regions in the left panel represent high density material, with molecular clouds being
represented by black. Electron densities smaller than $\log n_{e}=-4$ (white regions;
right panel) are located in both high (atomic and molecular clouds) and lower (bubbles) density regions.}
\label{fig1}
\end{figure*}

For our analysis we chose pulsars from the ATNF catalog \citep{mhth2005} with independent distance
estimates, i.e., estimates without using NE2001 model (Cordes \& Lazio 2003; CL03). These estimates resulted
from parallax measurements, absorption-line (\hi 21-cm or OH line) studies, physical associations or timing towards
pulsars in the galactic disk ($\left|z\right|< 200$ pc). Most parallax estimates were obtained from \citet{vlm2010},
which are updated by Chatterjee, S.\footnote{\url"www.astro.cornell.edu/~shami/psrvlb/parallax.html"}.

For our study we augmented the ATNF catalog by including distance estimates from absorption studies wherever available
from the literature, including two measurements from associations (Frail et al, 1996; Saravanan et al. 1996; Johnston et
al. 2003;  Minter et al 2008; Weisberg et al 2008; see references therein). To quantify errors in
distance estimates, we compute the fractional error involved in parallax and absorption studies 
from the difference between the upper and lower estimates. Firm error estimates  are not available for several 
pulsars in the ATNF catalog, whereby the fractional error was computed from the difference between 
the distance estimate and the best  estimate available from NE2001 model  (parameter DIST1 in ATNF 
catalog). Errors in distance estimates using  parallax measurements range between $2-35\%$. The median error in distance
estimate in our overall sample is $18\%$, a significant improvement over estimates 
using NE2001 model (CL03). 

A total of 122 pulsars with reliable distance estimates were thus selected, $33$ of which have 
$\left|z\right|< 200$~pc. Nine pulsars were removed from this sample as they are located within the vicinity
of the Sun, and measurements of electron density towards those sightlines are significantly affected by local
structures, such as the Local Bubble, Gum nebula and Loop~I (e.g., CL03). The histogram of the
computed $\log \langle n_{e} \rangle$ derived from the sample DMs and the best-fit gaussian curve
are shown in Figure~\ref{observ} (black lines). The fit is centered at $\log\langle n_{e}
\rangle=-1.47\pm0.02$ and has a dispersion  of $\sigma=0.17\pm 0.02$, with a $\chi^2/\mbox{dof}=1.62$.
These values compare well with those derived directly from our pulsar sample data: $\log\langle n_{e}
\rangle\sim 1.42$ and $\sigma\sim 0.23$.

\section{Simulation Results}

The simulated supernova-driven interstellar medium is characterized by several evolutionary phases that have already
been described in previous works: (i) domination of initial conditions being only wiped out after some 80 Myr of
evolution, (ii) the full establishment of the continuous disk-halo-disk circulation (also known as the Galactic
fountain), and (iii) the dynamical equilibrium (occurring at 200-220 Myr of evolution) in a statistical
sense that determines the dynamics of the interstellar medium in the Galactic disk and its interaction with
the halo. As a result of this evolutionary path driven mainly by supervovae, the ISM becomes frothy and
turbulent with a mean Mach number of 3 (AB0507). Low temperature gas (high density) is concentrated
into filamentary structures and molecular clouds (black regions in the left panel of Figure~\ref{fig1}, showing the
Galactic midplane density at 400 Myr of evolution), while hot gas (low density regions in the same panel) is
concentrated into bubbles and superbubbles that dominate the landscape. \\
\begin{figure}
\centering
\includegraphics[width=0.92\hsize,angle=0]{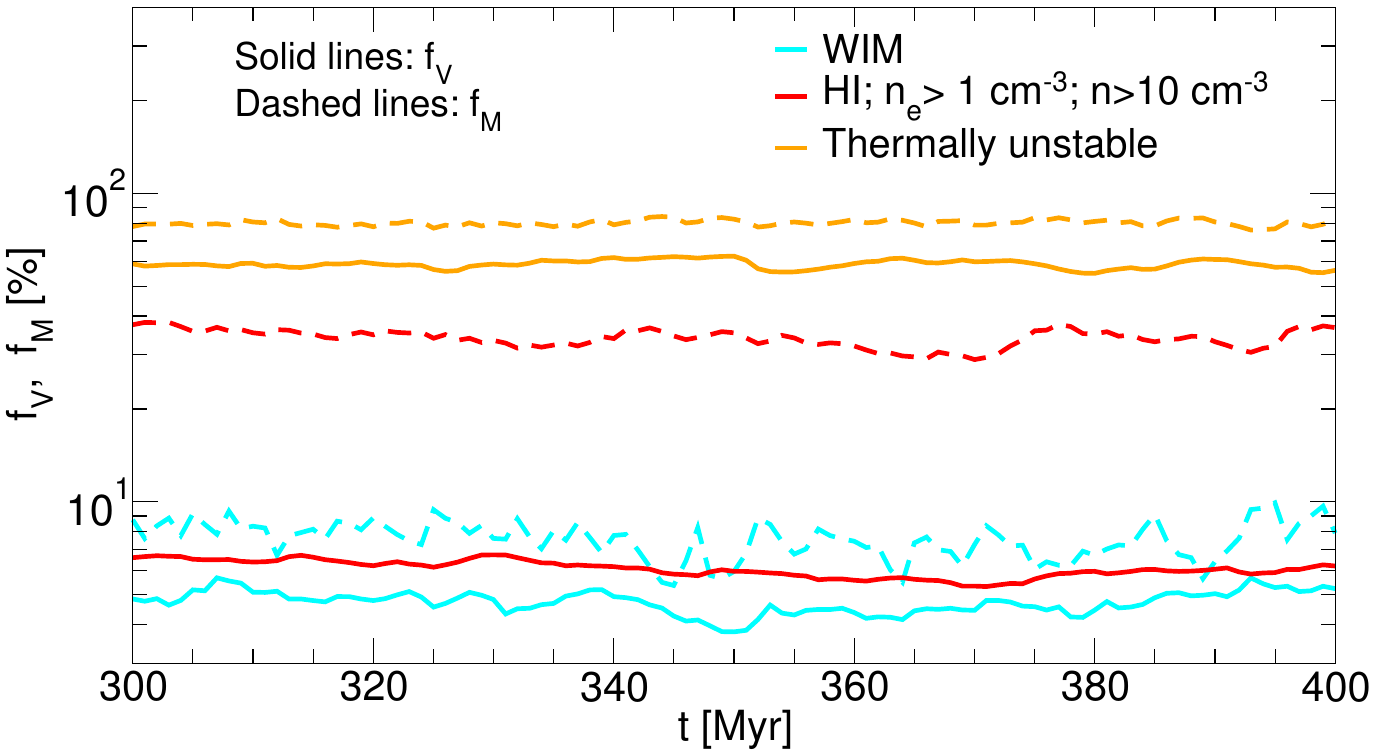}\\
\includegraphics[width=0.92\hsize,angle=0]{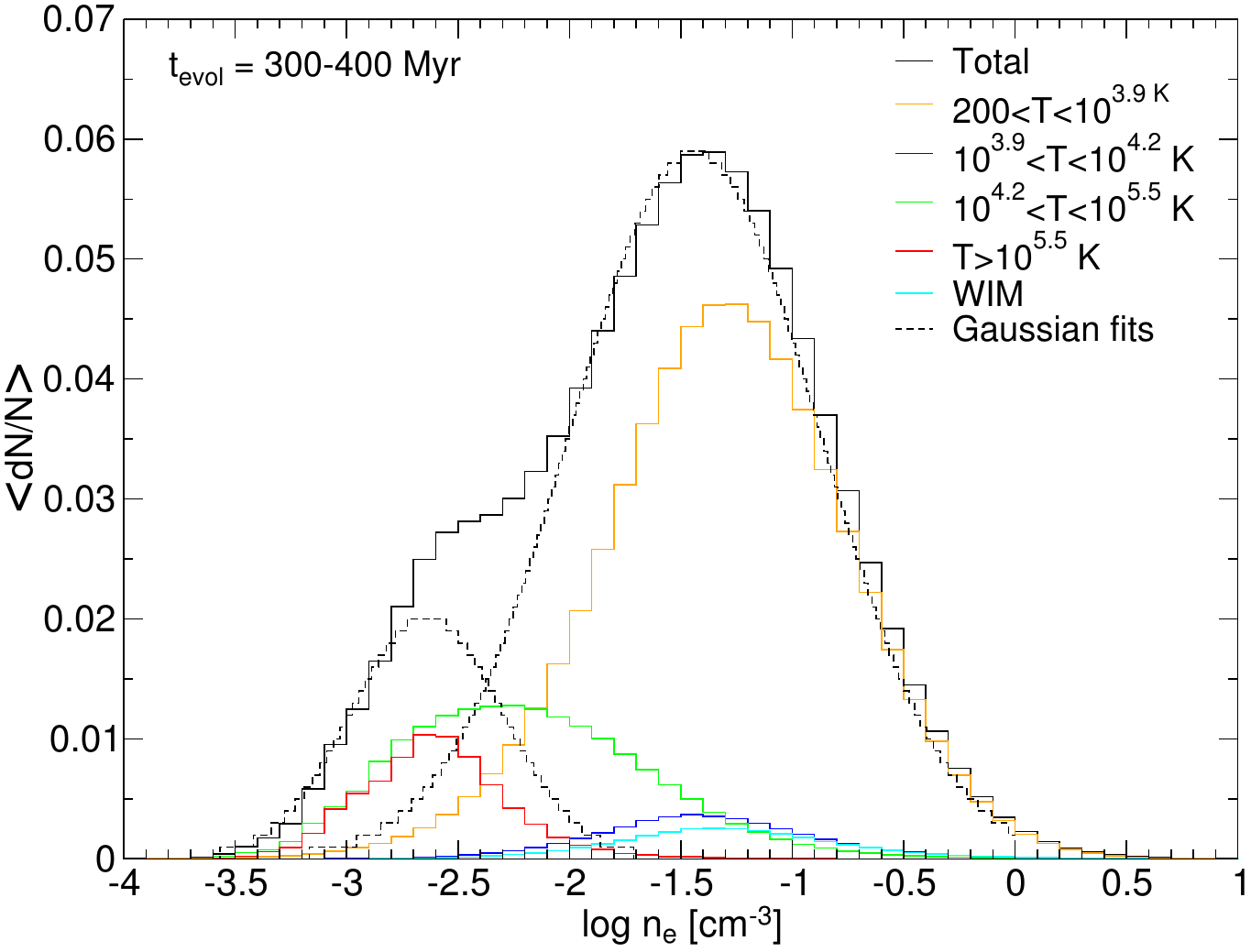}
\caption{\emph{Top panel:} Time evolution between 300 and 400 Myr of the electrons volume (solid lines) and mass (dashed
lines) occupation fractions in the simulated disk for thermally unstable regime (orange), gas in shells and
filamentary structures (red), and warm ionized medium (WIM; cyan). \emph{Bottom panel:} Time averaged histogram of the
total (black ), and the two gaussian fits (dashed black lines) of the electron density in the simulated disk over the
evolution time 300-400 Myr. The electron density in different temperature regimes are displayed.}
\label{avepdf}
\end{figure}
\indent The turbulent nature of the simulated ISM, the ongoing physical processes and driving mechanisms
(e.g., SNe and stellar winds) affect the electron distribution, which follows the topology of the
medium. Most of the electrons are distributed into the thermally unstable regime
($200<\mbox{T}<10^{3.9}$ K) with a volume ($f_v$) and mass ($f_M$) filling fractions of $56-60\%$
and $77-80\%$, respectively (Figure~\ref{avepdf}: top panel); $32-40\%$ of the electron mass
is locked in filamentary structures and shells, having $f_{v}=6-7\%$. The electron
distribution in warm ionized medium has $f_v\simeq 4.9-6\%$ and $f_M=7-10\%$. These are
time-dependent variations whose time average (over a period of 100 Myr using 1001 disk snapshots
taken at every 0.1 Myr) histogram of $\log n_{e}$ (Figure~\ref{avepdf}: bottom panel), is fitted
with a composition of two gaussians centered at $\log n_{e}=-1.437$ (for T$<10^{4.2}$ K gas) and
$\log n_{e}=-2.64$ (for T$>10^{4.2}$ K gas), respectively, with the latter representing mostly photoionized
regions.

Further insight into the simulated electron distribution can be obtained directly from DMs
along LOS, with length $d$, crossing the simulated ISM, allowing the determination of the mean
electron density. The LOS have increasing lengths (with a step length of 10 pc) up to 1 kpc from a vantage
point located at $(x,y,z)=(0,0,0)$ (left bottom corner of the electron density maps in Figure~\ref{fig1}) for
$\left|z\right|\leq 200$ pc and spanning $90\degr$ (with a $1\degr$ separation) when projected onto the
Galactic midplane.
\begin{figure}
\centering
\includegraphics[width=0.52\hsize,angle=-90]{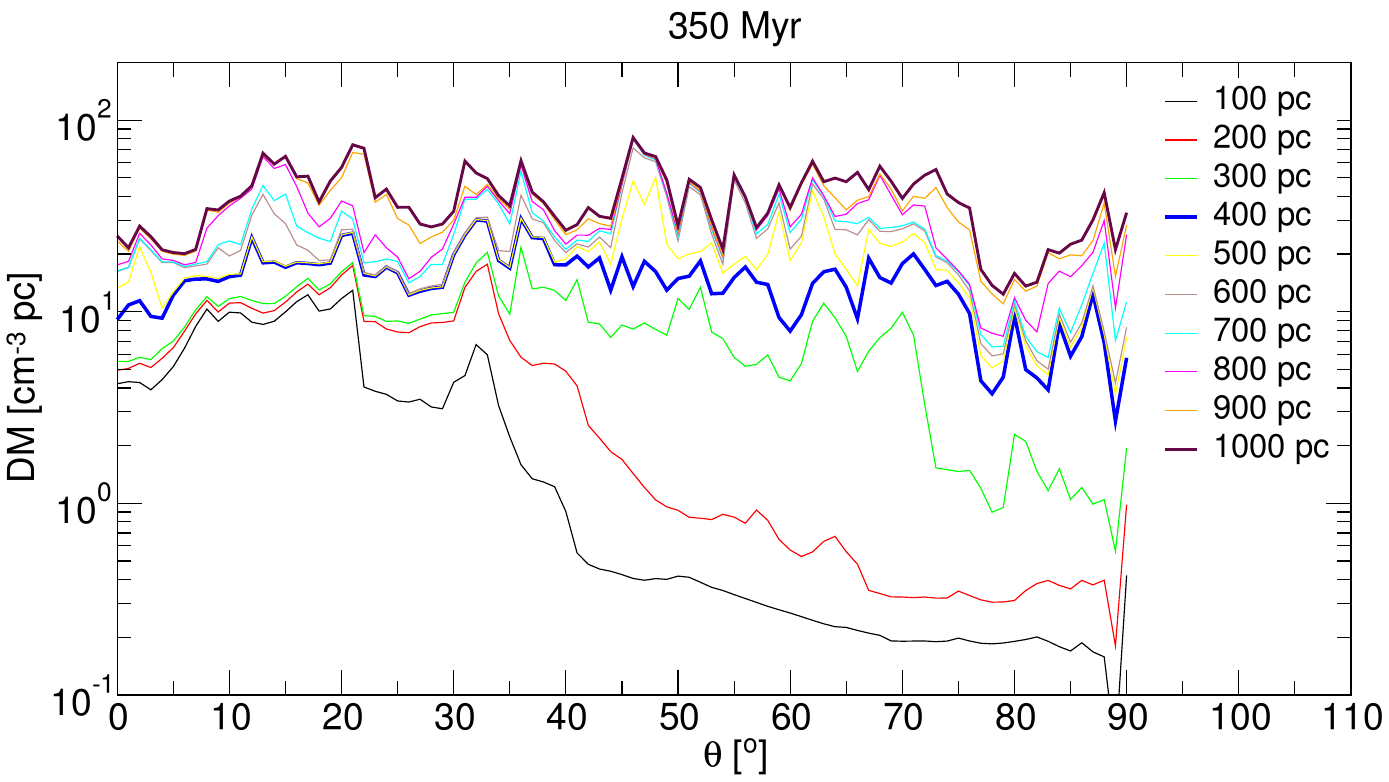}\\
\includegraphics[width=0.52\hsize,angle=-90]{plots/mavillez_fig4a.pdf}\\
\includegraphics[width=0.52\hsize,angle=-90]{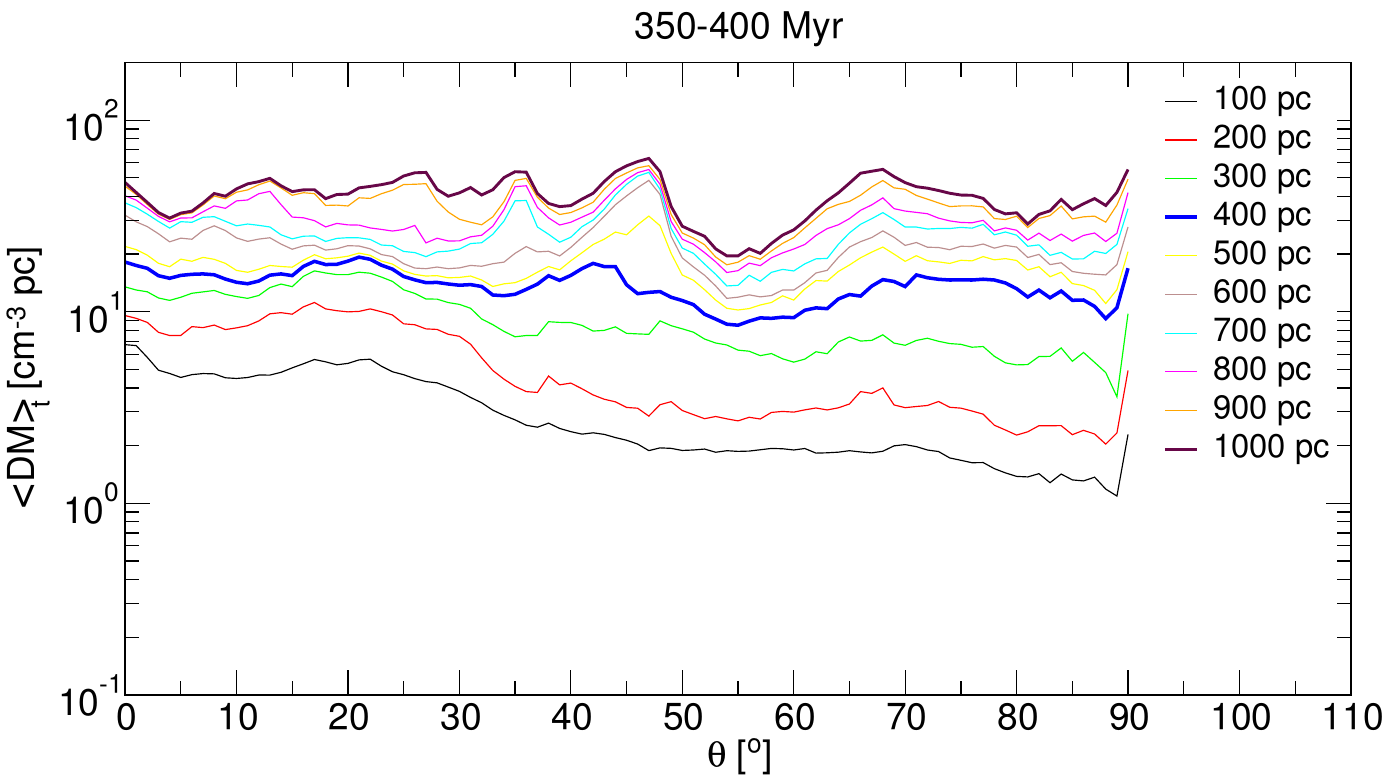}
\caption{Angular distribution of the DMs along LOS with lengths varying between 100 and 1000
pc at 350 and 400 Myr (top two panels) and the time averaged $\langle DM\rangle_{t}$ (bottom panel) over the
evolution time 350-400 Myr).}
\label{dmangle}
\end{figure}

Figure~\ref{dmangle} shows the DM along the LOS with lengths varying between 100 and 1000 pc, with a step
length of 100 pc at 350 and 400 Myr (top two panels) and its time average (bottom panel) over 501 disk
snapshots taken every 0.1 Myr for $350-400$ Myr of evolution. Due to the ongoing turbulent processes in the
Galactic disk the DM has variability along and between the lines of sight (Figure~\ref{dmangle}: top two
panels). The longitudinal variation is due to the inhomogeineity of the ISM (e.g., bubbles and superbubbles)
as well as due to the turbulent nature of the regions crossed by the LOS. These variations with distance and
among LOS are still present in the time averaged DM, although the longitudinal variations are smoothed in the
averaging process. However, a record of any event that prevailed over a large period of time (e.g., a
superbubble whose contribution to the DM is small as in the case of the observed hump in the DM between
48\degr and 65\degr) is kept.
\begin{figure}
\centering
\includegraphics[width=0.92\hsize,angle=0]{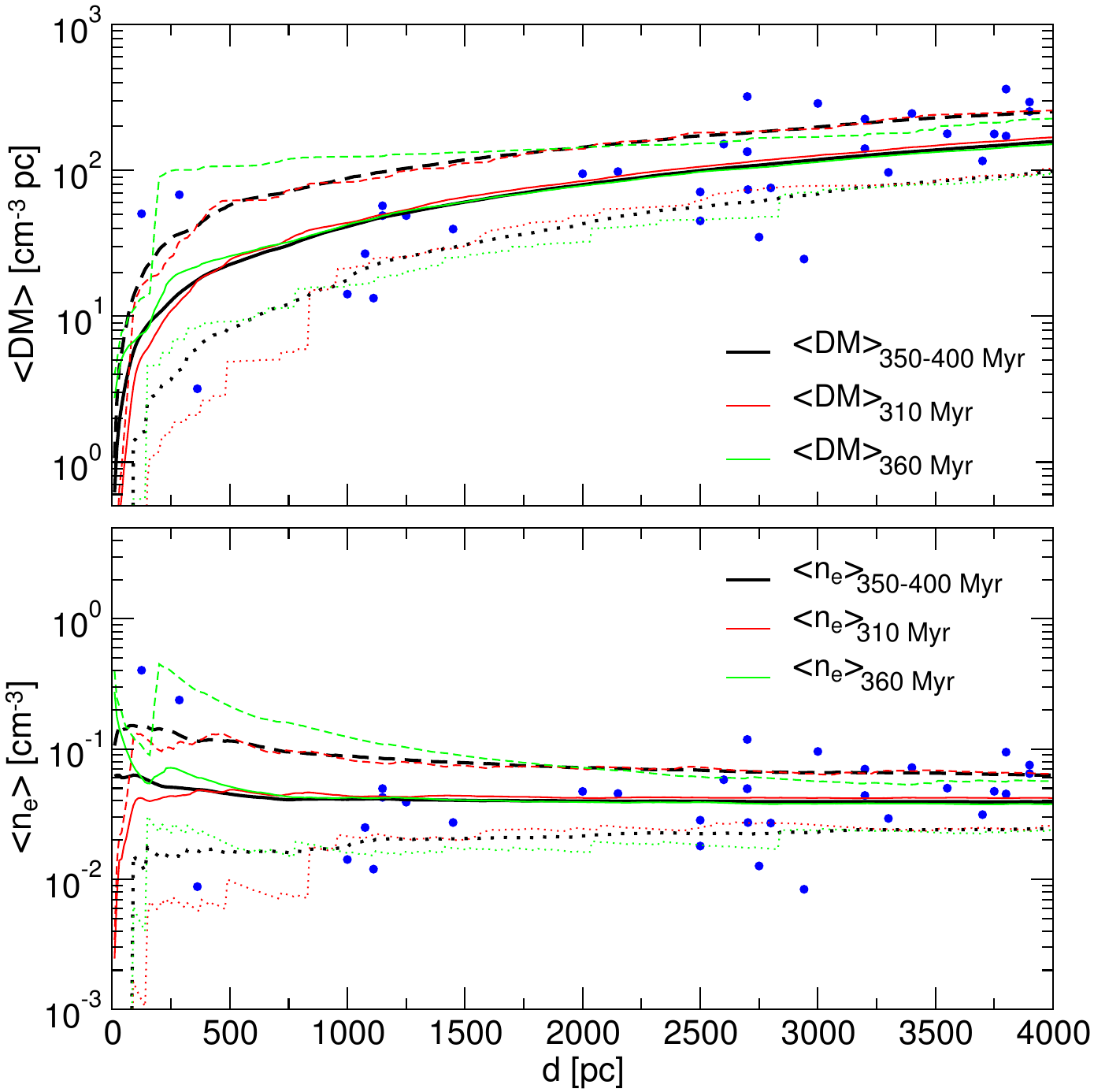}
\caption{Minimum (dotted line), mean (solid line) and maximum (dashed line) of the averaged DM (top panel) and
$n_{e}$ (bottom panel) for all LOS crossing the disk up to 4 kpc from the Sun at 310 Myr ($\langle
X\rangle_{310\mbox{ Myr}}$; red lines), 360 Myr ($\langle X\rangle_{360\mbox{ Myr}}$; green lines), and over
the period 350-400 Myr ($\langle X\rangle_{350-400\mbox{ Myr}}$; black lines). The blue circles represent
the DMs (top) and $\langle n_{e}\rangle$ (bottom) derived from pulsars observations.}
\label{taver}
\end{figure}

We further compare the DMs and $\langle n_{e}\rangle$ derived from pulsars observations (\S3), with those
resulting from averaging over all the LOS crossing the simulated disk ($|z|<200$ pc) and up to 4 kpc from the
Sun at specific times (e.g., 310 and 360 Myr) and over the period 350-400 Myr. Top panel of Figure~\ref{taver}
displays the observationally derived DMs and the averaged DMs at 310 Myr ($\langle DM\rangle_{310\mbox{
Myr}}$; red lines), 360 Myr ($\langle DM\rangle_{360\mbox{ Myr}}$; green lines), and over the period 350-400
Myr ($\langle DM\rangle_{350-400\mbox{ Myr}}$; black lines).

Due to the limitation of the box size in our simulation, we calculated the DMs for distance $d>1$ kpc taking
advantage of the periodic boundary conditions used in the simulations. The DMs were calculated from different
vantage points in the midplane (including $(x=0,y=0,x=0)$ and between 500 and 1400 pc from it) using LOS
crossing the disk volume at different angles, up to 4 kpc from the vantage point and up to $z=\pm200 $ pc.
This procedure, although inevitable here, may introduce artefacts in the simulation results, but we do not
expect that our results are strongly affected though, because the simulation shows a certain pattern
repetition in the ISM density and temperature distributions on scales of the order of a few correlation
lengths (that is, a few times 75 pc, according to AB0507).

The averaged $\langle DM\rangle_{350-400\mbox{ Myr}}$, after a steep growth in the first 100 pc, reaches a
smooth increase for $d>500$ pc, with the deviations of the minimum and maximum regarding the mean
becoming constant over distance - consistent with the DMs variations in $d$ observed in
Figure~\ref{dmangle}. The averaged DMs at 310 and 360 Myr vary widely, but with values within the pulsar
derived DMs. Therefore, the terms minimum and maximum of the $\langle DM\rangle_{350-400\mbox{ Myr}}$ should
not be taken as strict upper and lower values for all times, as they only represent an average over a
specific time window of the simulation.

After the large variation in the first few hundred parsecs, the mean of the simulated $\langle
n_{e}\rangle$, for the cases shown in the bottom panel of Figure~\ref{taver}, reach a constant value
($0.04\pm 0.002$ cm$^{-3}$) and dispersion over large distances. This reflects the clumpy nature of the
electrons in the turbulent ISM as would be expected if they are predominantly found in filaments and shells
(Berkhuijsen \& M\"uller 2008). The histogram of the $\log \langle n_{e}\rangle_{350-400\mbox{ Myr}}$ and
it's best Gaussian fit are displayed in Figure~\ref{observ} by solid and dashed red lines, respectively.
The fit is centered at $\log\langle n_{e}\rangle=-1.4\pm 0.01$ with $\sigma=0.21\pm0.01$. In the simulated
disk, the clumpy material dominates the LOS and volume averaged PDFs which explains the similar means between
the electron density of the T$<10^{4.2}$ K regime (Figure~\ref{avepdf}) and the time averaged $\langle
n_{e}\rangle_t$ (Figure~\ref{observ}; red lines). These fit parameters are similar to those
displayed in Figure~\ref{observ} (black lines) for observationally derived $\langle
n_{e}\rangle$. This similarity is indicative of the smoothing out of low and high density
regions along the lines of sight, thus, not appearing in the PDFs, and leading to reduced Gaussian
dispersion.

\section{Discussion and Final Remarks}

In this letter we discuss the electron density distribution in the Galactic disk using the first to date 3D high
resolution NEI simulations of the ISM. The simulations trace the dynamical and thermal evolution of the interstellar
gas, calculating on-the-spot the ionization structure, electron distribution and cooling function of
the gas at each cell of the grid into which the computational domain is discretized. We did
not take into account stellar ionizing photons and their transport, but an averaged diffuse photon
field, which should be a reasonable approximation, when describing the mesoscale ISM. The NEI
structure modifies the cooling function, which in turn enhances the number of free electrons. 

Both the simulated DMs and electron density are consistent with the observations: (i)nmost DMs lie within the
maximum and minimum observed values, (ii) $\log \langle n_{e} \rangle$ is consistent with a gaussian
distribution, (iii) the mean electron density is $0.04\pm0.01$ cm$^{-3}$, and (iv) the volume filling
fraction of the WIM is bracketed between 4.9 and 6\%. These results lye in between the estimates by Gaensler
et al. (2008) and Berkuijsen \& M\"uller (2008) for the electron density in the disk
and WIM. Furthermore, the observed Gaussian distribution is present, irrespective of whether the ISM is
isothermal (see discussion in Berkuijsen \& Fletcher (2008) and refereces therein) or not. This can be
explained in two complementary ways: a fully developed turbulent system can be considered as a large number of
independent random variables, with the logarithm of each having a certain distribution approaching a Gaussian
as the number of variables goes to infinity (Central Limit Theorem), being the case of supernova-driven ISM
simulations reaching a statistical equilibrium (see, e.g., V\'azquez-Semadeni \& Garcia 2001). Or,
alternatively, we can use the principle of maximum entropy. The
information entropy of a continuous random variable $X$ with probability density function $p(x)$ is defined as
$H(X)=\int_{-\infty}^{\infty} p(x) \log p(x) dx$. The lognormal distribution of $X$ maximizes $H(X)$,
implying the least prior knowledge of the system (Sveshnikov, 1968). This is exactly what is expected in
homogeneous and isotropic turbulence with a large number of independent random variables. We study
deviations from lognormal distributions in a forthcoming paper.

\paragraph*{Acknowledgements}
We thank the anonymous referee for the detailed report and valuable suggestions that allowed us to improved
this letter. The Milipeia Supercomputer (Univ. of Coimbra) and the ISM-cluster (Univ. of \'Evora) were heavily
used for the calculations. This research is supported by the FCT project PTDC/CTE-AST/70877/2006.

\bibliographystyle{mn2e}

\label{lastpage}
\end{document}